\documentclass{jetpl}
\twocolumn

\usepackage{caption}
\usepackage{graphicx}
\usepackage{cite}
\usepackage[dvips]{graphicx}
\DeclareGraphicsExtensions{.pdf,.png,.jpg,.eps,.ps}
\RequirePackage{caption}
\DeclareCaptionLabelSeparator{point}{. }
\title{Sources of primary cosmic rays}
\rtitle{Sources of primary cosmic rays}
\sodtitle{Sources of primary cosmic rays}

\author{S.\,E.\,Pyatovsky \/\thanks{ mail: vgsep@ya.ru\\
ORCID: 0000-0003-2565-1670}}
\rauthor{S.\,E.\,Pyatovsky}
\sodauthor{Pyatovsky}

\address{Lebedev Physical Institute of the Russian Academy of Sciences}

\dates{May 30th, 2023}

\abstract
{In the paper, a comparative primary cosmic rays (PCR) comparative analysis by $E_0$ and the spectra of variable stars by periods is carried out in order to establish the causes of irregularities in the spectrum of PCR by $E_0$. The study was performed using the public database of the \mbox{KASCADE-Grande} experiment and GCVS and ZTF variable star catalogues. It has been suggested that the acceleration of PCR to high and super-high energies occurs not only on the shock waves of supernovae, but also in bursts of giants and super-giants. The relationship between the periods of variable stars and the maximum energy $E_0$ of the nuclei of PCRs generated by these types of stars is shown. Irregularities in the PCR spectrum by $E_0$ are associated with the transition from one dominant stars type to another as $E_0$ increases. The knee in the PCR spectrum at $E_0~=~3-5~PeV$ is associated with a decrease in the contribution of SRB variability stars and a further increase in the contribution of Mira variable stars to the PCR flux. The bump in the PCR spectrum with a maximum at $E_0~=~80~PeV$, established in the \mbox{KASCADE-Grande} experiment, is formed by giant stars and super-giants of the Mira and SRC variability.}


\begin{document}

\maketitle
{\bf Introduction.}
The reasons for and type of irregularities in the PCR spectrum by $E_0$ remain the subject of scientific discussions. The issues are discussed of the so-called knees localization in the spectrum, the "sharpness" of knees, at what energies knees are observed in the spectra of light and heavy nuclei in the mass composition of PCR, and others. Attention is drawn to the issues of localization and the source of the bump in the $E_0$ spectrum of PCR at energy of about $100~PeV$.

\begin{figure}[h]
\centering
\includegraphics[width=\linewidth]{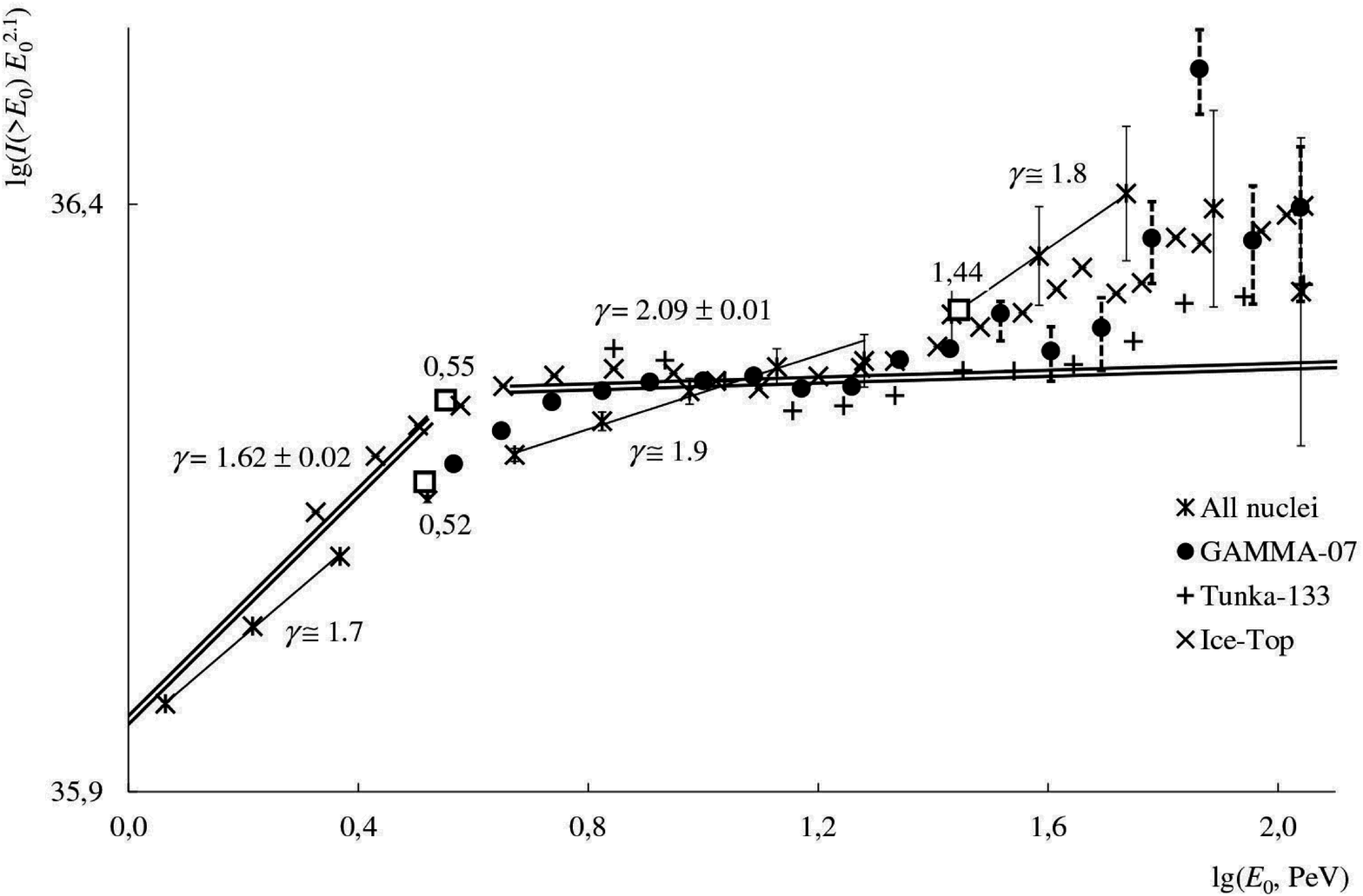}
\caption{Integral spectra by $E_0$ obtained in the \mbox{KASCADE-Grande} and \mbox{GAMMA-07} experiments. The bump maxima in the spectra obtained in the \mbox{KASCADE-Grande} and \mbox{GAMMA-07} experiments correspond to $E_0~=~50-80~PeV$, which is higher than the knee energy of the heaviest nuclei in the PCR mass composition, as shown in~\cite{1}. The outlier point (circled) is observed in the PCR spectrum by $E_0$ confirmed in the \mbox{GAMMA-07} experiment. The bump with a maximum at $E_0~\cong~80~PeV$ was confirmed by the \mbox{KASCADE-Grande} experiment.}
\label{fig1}
\end{figure}

Analysis of irregularities in the PCR spectrum at $E_0~=~1-100~PeV$ was carried out, in particular, in~\cite{1}. Figure~\ref{fig1} shows the results of the \mbox{KASCADE-Grande}~\cite{2}, Tunka and Ice-Top experiments on measuring the energy spectrum of PCR. Special attention should be paid to the results of the GAMMA (\mbox{GAMMA-07}) experiments~\cite{3} (Armenia, Mount Aragats) and "Hadron" (Tien Shan high-mountain scientific station), $700~g/cm^2$ of atmospheric depth, in which at $E_0~\cong~70-100~PeV$ intensity peak was registered, shown in Figure~\ref{fig1}, which significantly exceeds the data of other experiments. The nature of this irregularity has not been established, and other experiments do not report the presence of a similar peak. It is also unusual that this peak was not observed in the exposures of the GAMMA experiment itself (Armenia), but for other time periods, for example, \mbox{GAMMA-06}, -08, etc. However, it is quite possible that this peak is not a methodological error in the processing of experimental data.

It was shown in~\cite{1,4} that the irregularities in the spectrum of PCR by $E_0$ following the knee at $E_0~=~3-5~PeV$ are due to the departure of PCR mass composition nuclei starting with protons. Using the “min-max of EAS age” method~\cite{1}, based on the sufficiently large statistics of EAS experimental characteristics obtained in particular in the \mbox{KASCADE-Grande} experiment, it was shown that at $E_0~=~2-35~PeV$, the mass composition of PCR nuclei remains mixed and corresponds to the CNO group. However, the knee in the nuclei spectrum of the PCR mass composition of the heaviest group is localized before the bump observed at $E_0~=~50-100~PeV$, which indicates that the bump in the spectrum of PCR at $E_0~=~50-100~PeV$ is formed by other sources of nuclei and with different acceleration features.

{\bf 1 Experimental data for bump analysis at \boldmath$E_0~=~50-100~PeV$.}
The bump analysis of the PCR spectrum by $E_0$ was performed using the data of the \mbox{KASCADE-Grande} experiment~\cite{2,4}, the database of which contains characteristics of more than 150 million EASs, including EAS global registration time in the range from 846,252,788 to 1,071,878,399 seconds with the count starting from 01/01/1970.

The characteristic of this irregularity (bump) under study is the slope index $\gamma$ of the PCR spectrum by $E_0$. To estimate the change in $\gamma$, a range of $E_0~=~20-75~PeV$ was chosen located after the knee in the group of the heaviest nuclei in the PCR mass composition of and up to the bump maximum at $E_0~=~80~PeV$ established by the \mbox{KASCADE-Grande} collaboration and confirmed in the GAMMA and Hadron experiments. The study of the change in $\gamma$ was performed with a lag of 10 days, which provided statistics for each sample of $\cong$~1~million events.

\begin{figure}[h]
\centering
\includegraphics[width=\linewidth]{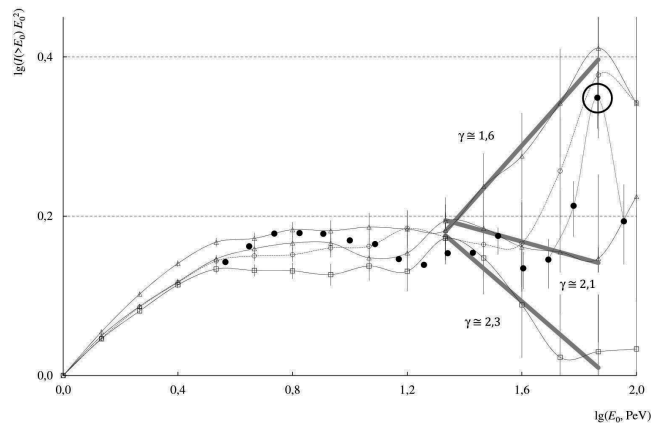}
\caption{PCR spectrum by $E_0$ obtained in the \mbox{GAMMA-07} experiment (black circles, outliers circled) as compared with the data of the \mbox{KASCADE-Grande} experiment for different time intervals (empty markers). Gray lines are regressions in the range of $E_0~=~20-75~PeV$ (up to the bump maximum at $E_0~=~80~PeV$, established in the \mbox{KASCADE-Grande} experiment).}
\label{fig2}
\end{figure}

Examples of spectra for 10-day samples are shown in Figure~\ref{fig2}. Figure~\ref{fig2} shows spectra with slope indices $\gamma$ near the bump at $E_0~=~80~PeV$ from the minimum values $\gamma~=~1.60~\pm~0.02$ to the maximum values $\gamma~=~2.31~\pm~0.04$. The spectra constructed from samples from the \mbox{KASCADE-Grande} experimental data are compared with the data from the \mbox{GAMMA-07} experiment. Despite the fact that the $\gamma$ index averaged over the entire observational statistics was obtained with high accuracy, the values of $\gamma$ for different time intervals differ significantly. This change in $\gamma$ can be associated either with fluctuations in EAS characteristics, or with the intensity of PCR in the given range $E_0$. It also follows from Figure~\ref{fig2} that the outlier event recorded in the \mbox{GAMMA-07} experiment is not unique and has analogues in the events recorded in the \mbox{KASCADE-Grande} experiment.

Using the database of the \mbox{KASCADE-Grande} experiment, the values of the slope index $\gamma$ near the bump of the PCR spectrum were obtained in the range of $E_0~=~20-75~PeV$ for 248 time intervals.

\begin{figure}[h]
\centering
\includegraphics[width=\linewidth]{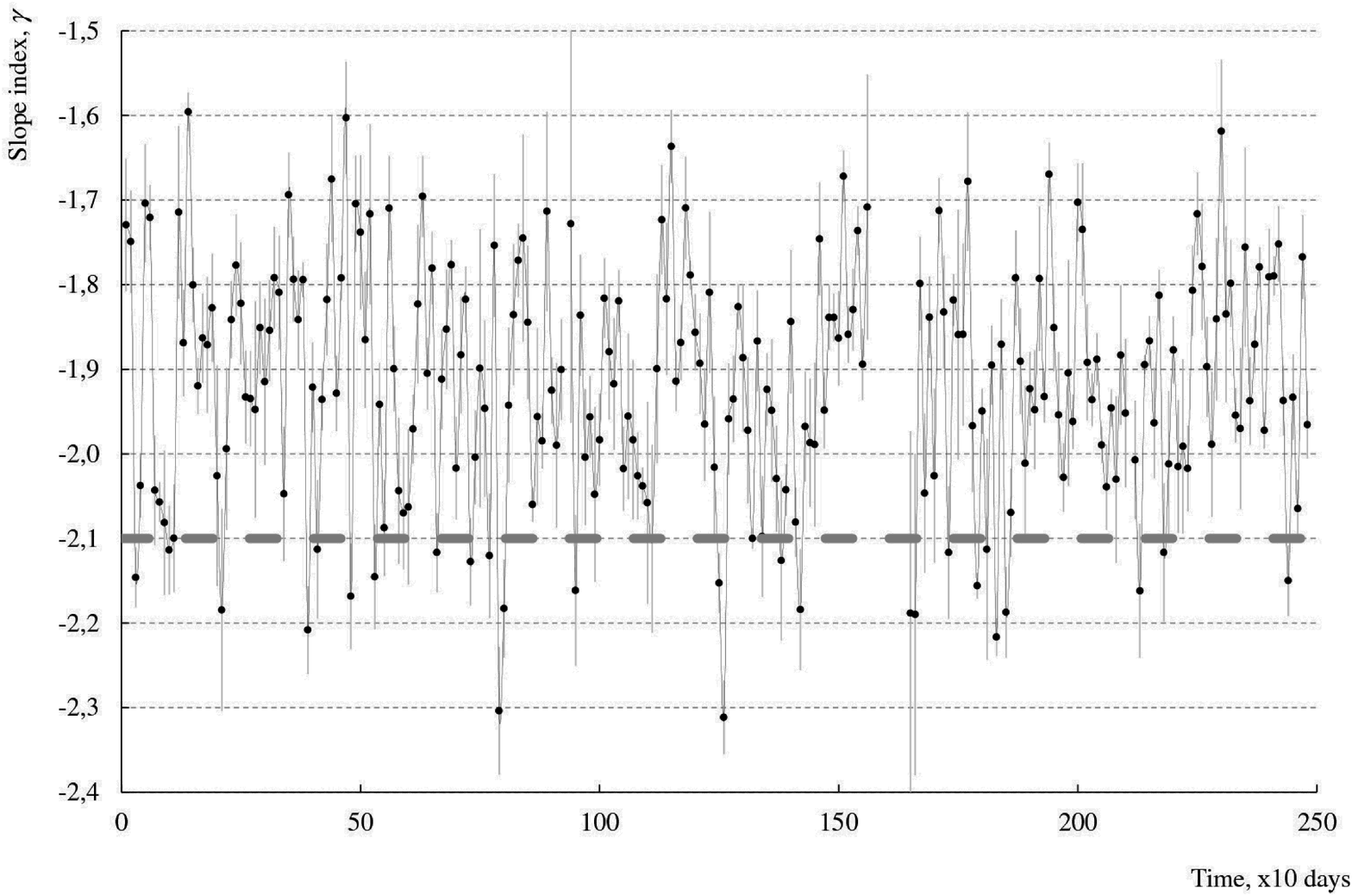}
\caption{Change in the PCR $\gamma$ spectrum index by $E_0$ over time. The horizontal dotted line corresponds to the value of $\gamma$ in the absence of a bump in the range of $E_0~=~20-100~PeV$. The values of $\gamma$ above this dotted line indicate the presence of a bump with a maximum at $E_0~=~80~PeV$.}
\label{fig3}
\end{figure}

{\bf 2 Spectral analysis of the change in the \boldmath$\gamma$ index.}
Figure~\ref{fig3} shows the change in the $\gamma$ index over time. The spectral analysis of the change in $\gamma$ was performed in order to identify possible maxima of the periods of change in $\gamma$ values. The spectral Fourier transform with the Hamming’s window was used for the analysis. It follows from Figure~\ref{fig3} that the change in $\gamma$ goes beyond the standard deviation, which allows us to assume the presence of regular PCR sources in the range $E_0~=~20-100~PeV$.

\begin{figure}[h]
\centering
\includegraphics[width=\linewidth]{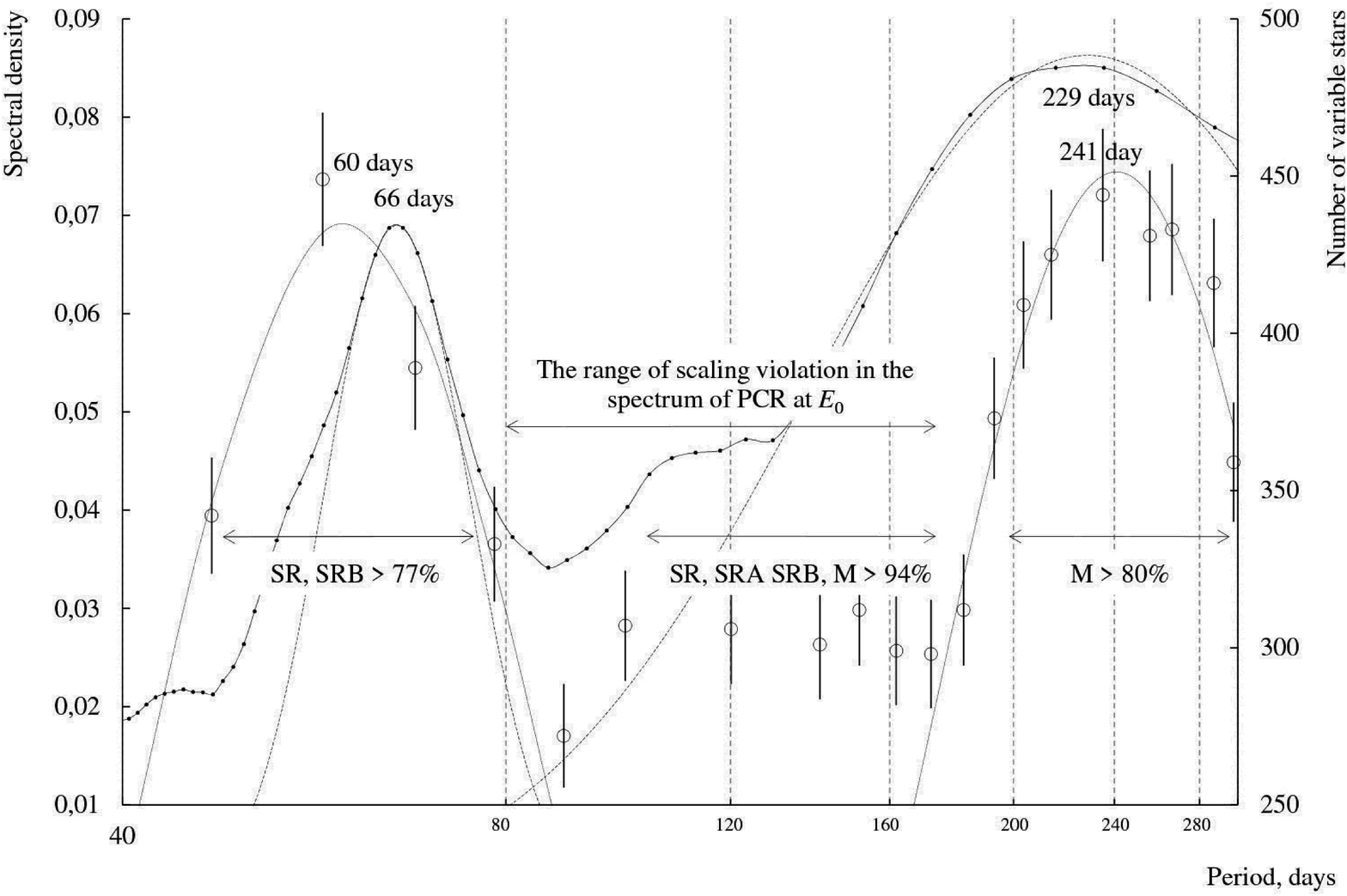}
\caption{Spectral density of the log-period of the PCR spectrum $\gamma$ index change by $E_0$ (solid line, main axis) in comparison with the periods of stars of various types of variability (dashed lines, additional axis): SR - semi-regular red giants and super-giants of intermediate or late spectral classes, SRA and SRB - semi-regular red giants of late M, C and S spectral classes, M - Miras, red giants at the final stages of stellar evolution with emission spectra of late classes.}
\label{fig4}
\end{figure}

The resulting spectral density of the log-period is shown in Figure~\ref{fig4}. The analysis made it possible to identify two maxima in the period of $\gamma$ change in the interval of $40-300$ days, equal to 66 and 229 days, corresponding to the maximum spectral density. The width of the spectral density peak characterizes the “locality” of the PCR source: the closer the peak is to the normal distribution, the more likely one PCR source dominates in peak formation. The wider peak is formed by superposition of PCR sources of the same type. In Figure~\ref{fig4}, peaks with maxima in periods equal to 66 and 229 days are described by normal distributions with $R_a^2~>~98\%$ (Table~\ref{tabl:2}).

To search for possible PCR sources in the range $E_0~=~20-100~PeV$, the catalogs of stellar objects "General Catalog of Variable Stars (GCVS)"~\cite{5} and "Zwicky Transient Facility Catalog (ZTF)"~\cite{6} were considered. More than 60,000 stars of more than 250 types are represented in GCVS with indication of periods, locations and other characteristics. Figure~\ref{fig4} shows that the first harmonic (66 days) corresponds mainly to stars with SR-type variability, while the second harmonic (229 days) is formed mainly by Miras. It also should be noted here that stars that are at the final stages of evolution usually have strong magnetic fields.

The region of transition from semi-regular giants to Miras (Figure~\ref{fig4}) is characterized by a local violation of scaling in the PCR spectrum at $E_0~=~3-20~PeV$~\cite{7}. There should be many local regions similar to those shown in Figure~4 where scaling violation occurs, in the PCR spectrum by $E_0$,~- scaling violation regions are associated with the transition from one dominant star type to another, and the degree of scaling violations manifestation is determined by energy distributions which are provided by the dominant type of stars of the considered variability.

\begin{figure*}[h]
\centering
\includegraphics[width=\linewidth]{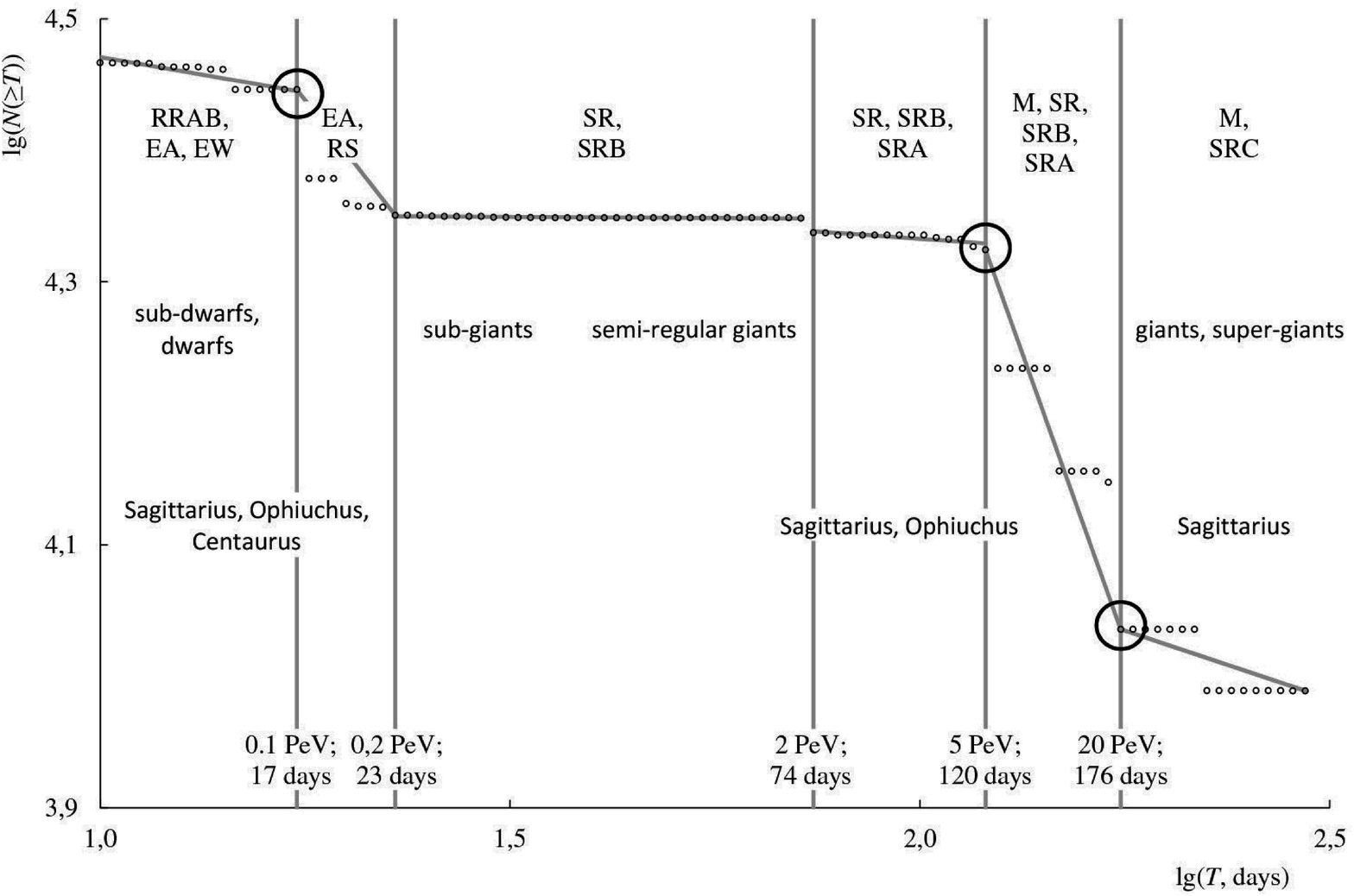}
\caption{Integral spectrum of variable stars over the log-period (empty circles). The CR acceleration is determined by "the type of variability of a star - the period of a considered star" factor. The dominant types of stars and the constellations where these stars are located are indicated for the indicated intervals by E0 and periods: EA - orange sub-giants of late evolution, EW - contact yellow dwarfs of spectral class F, RS - eruptive yellow-white dwarfs with a secondary component of a magnetically active sub-giant with spectra of Ca-II, H, and K in emission; SRC are semi-regular super-giants of the late M, C, and S spectral classes. Irregularities in the spectrum with assumed values of $E_0$ are circled.}
\label{fig5}
\end{figure*}

{\bf 3 Spectrum of stars by period.}
The integral spectrum of variable stars depending on the log-period is shown in Figure~\ref{fig5}, which shows the periods averaged over the types of variable stars. 252 types of stars have been considered, from white dwarfs of the ZZ Ceti type to super-giants of the recurrent nova type.

Let us assume that the more significant the irregularities in the spectrum by period of PCR sources are, the more significant are the irregularities in the spectrum of PCR by $E_0$ formed by these sources. The largest irregularities in the spectrum by period are indicated in Figure~\ref{fig5} as the known values of $E_0$: the period of 17 days corresponds to $E_0~=~0.1~PeV$ (red dwarfs region), 120 days corresponds to $E_0~=~5~PeV$ (the bump at $E_0~=~3-5~PeV$ in the PCR spectrum) and 176 days,~- $E_0~=~20~PeV$ (beginning of the bump with a maximum at $E_0~=~80~PeV$). The acceleration of PCR up to $E_0~=~0.1~PeV$ in flares of red dwarfs was established in the works by Yu.~I.~Stozhkov~\cite{8}.

Figure~\ref{fig5} shows that PCR sources of low energies $E_0~<~0.1~PeV$ are dwarfs localized mainly in the constellations of Sagittarius, Ophiuchus and Centaurus, with medium energies $E_0~=~0.2-2~PeV$,~- sub-giants and giants from the constellations of Sagittarius and Ophiuchi, and high energies $E_0~>~5~PeV$,~- giants and super-giants from the constellation Sagittarius. It should be noted here that the constellation determines the direction of arrival of PCR of given energies, but not the association with a given group of stars.

From the analysis of the data presented in Figure~\ref{fig5}, a regression was obtained that determines the relationship between the average period of a given type of PCR source star and the maximum $E_0$:

\begin{multline}
\mathrm{lg}(T, \mathrm{days})~=~(0.45~\pm~0.05)\mathrm{lg}(E_0, \mathrm{PeV})~+\\
+~(1.71~\pm~0.05)
\label{multline:1}
\end{multline}

Regression (1) is derived from the three hypothesized points shown in Figure~\ref{fig5} (circled). According to formula (1), the lower boundary of the sub-giant region for a period of 23 days is $E_0~=~0.2~PeV$, the region of the beginning of the first bump in the PCR spectrum by $E_0$ is a period of 74 days and $E_0~=~2-3~PeV$. It is also possible to estimate the maximum PCR energy. According to the GCVS catalog~\cite{5}, the maximum recorded period is 29,000 days (80 years) for recurrent novae of NR type variability. It follows from formula (1) that the period of 80 years corresponds to the maximum $E_0~\cong~1-2~ZeV$ registered in CR.

A few more points can be added to the data shown in Figure~\ref{fig5}, namely, expert estimates given in Table~\ref{tabl:1}.

\setcounter{table}{0}
\renewcommand\thetable{\arabic{table}}
\begin{table*}[htbp]
\caption{Dependence of the log-period of variable stars on the maximum energy $E_0$ in the PCR spectrum, which these sources provide.}
  \centering
    \begin{tabular}[t]{|c|c|l|}
\hline    
lg$(E_0,~PeV)$ & lg$(T,~days)$ & \multicolumn{1}{c|}{Notes} \\
\hline
-6.50 & -0.93 & H spectrum of the Sun \\        
-1.82 & 1.06 & Early break in the spectrum of PCRs (37th RCRC 2022) \\        
-1.00 & 1.23 & Red dwarfs (Yu.~I.~Stozhkov) \\        
0.48 & 1.87 & Beginning of the first bump in the PCR spectrum \\        
0.70 & 2.08 & The end of the first bump in the PCR spectrum \\        
1.00 & 2.02 & "Vela" local source model (A.~D.~Erlykin, V.~P.~Pavluchenko) \\        
1.30 & 2.25 & Beginning of the second bump in the PCR spectrum \\        
6.70 & 4.46 & Maximum registered energy of PCR nuclei \\
\hline
    \end{tabular}
\label{tabl:1}    
\end{table*}

The regression built according to the data of Table~\ref{tabl:1} is similar to (1) within the limits of errors:

\begin{multline}
\mathrm{lg}(T, \mathrm{days})~=~(0.41~\pm~0.01)\mathrm{lg}(E_0, \mathrm{PeV})~+\\
+~(1.71~\pm~0.03), R_a^2~\sim~1
\label{multline:2}
\end{multline}

{\bf 4 Types of variable stars and the PCR spectrum by \boldmath$E_0$.}
It follows from Figures~\ref{fig4} and \ref{fig5} that semi-regular giants and Miras make up the main stellar population providing PCR sources at $E_0~=~1-100~PeV$. Let us apply the “main array” method and consider the formation by stars of SR, SRA, SRB, and M type variabilities of the PCR spectrum for given $E_0$.

\begin{figure*}[h]
\centering
\begin{tabular}{cc}
\includegraphics[scale=0.15]{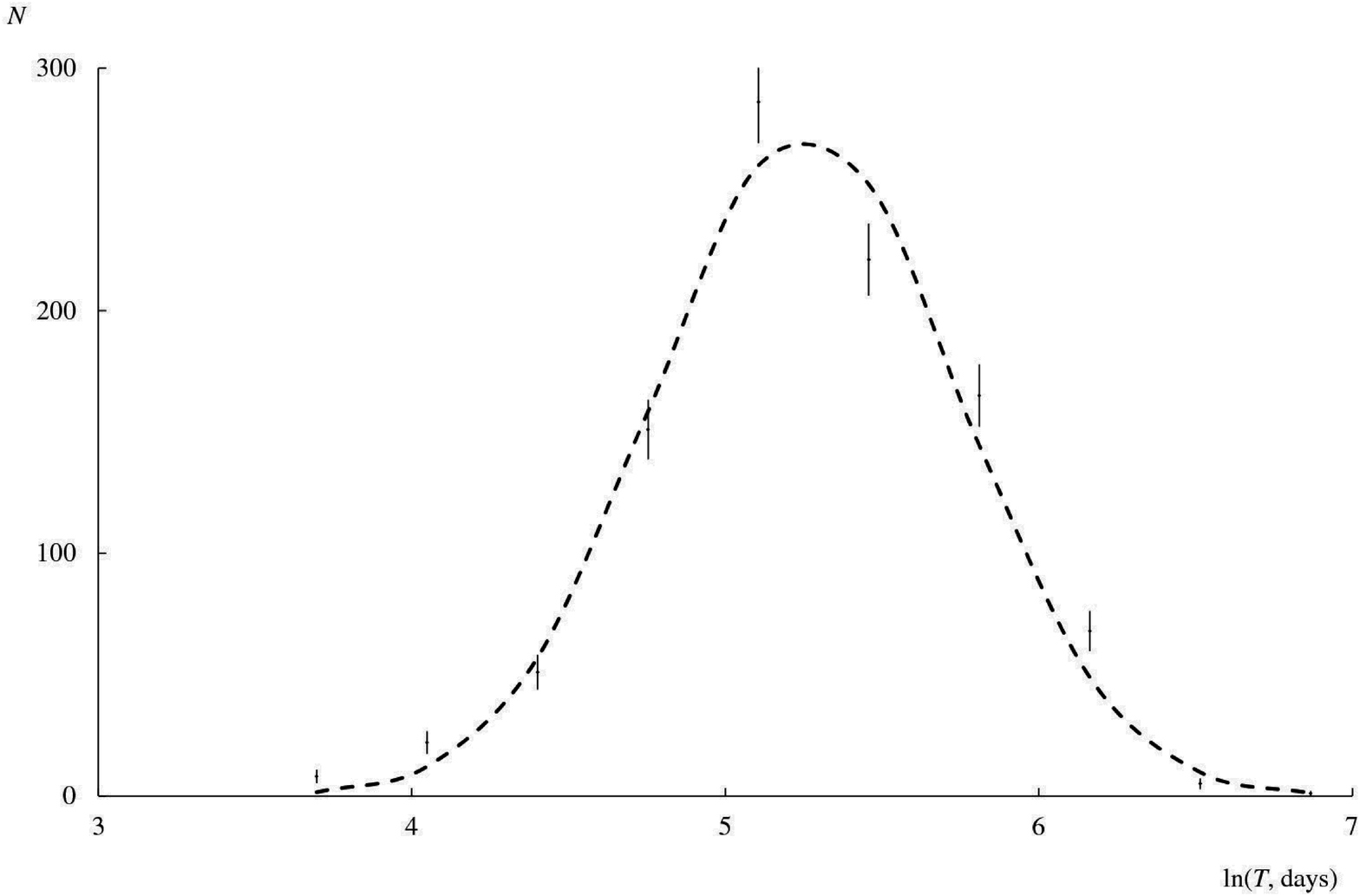}
&
\includegraphics[scale=0.15]{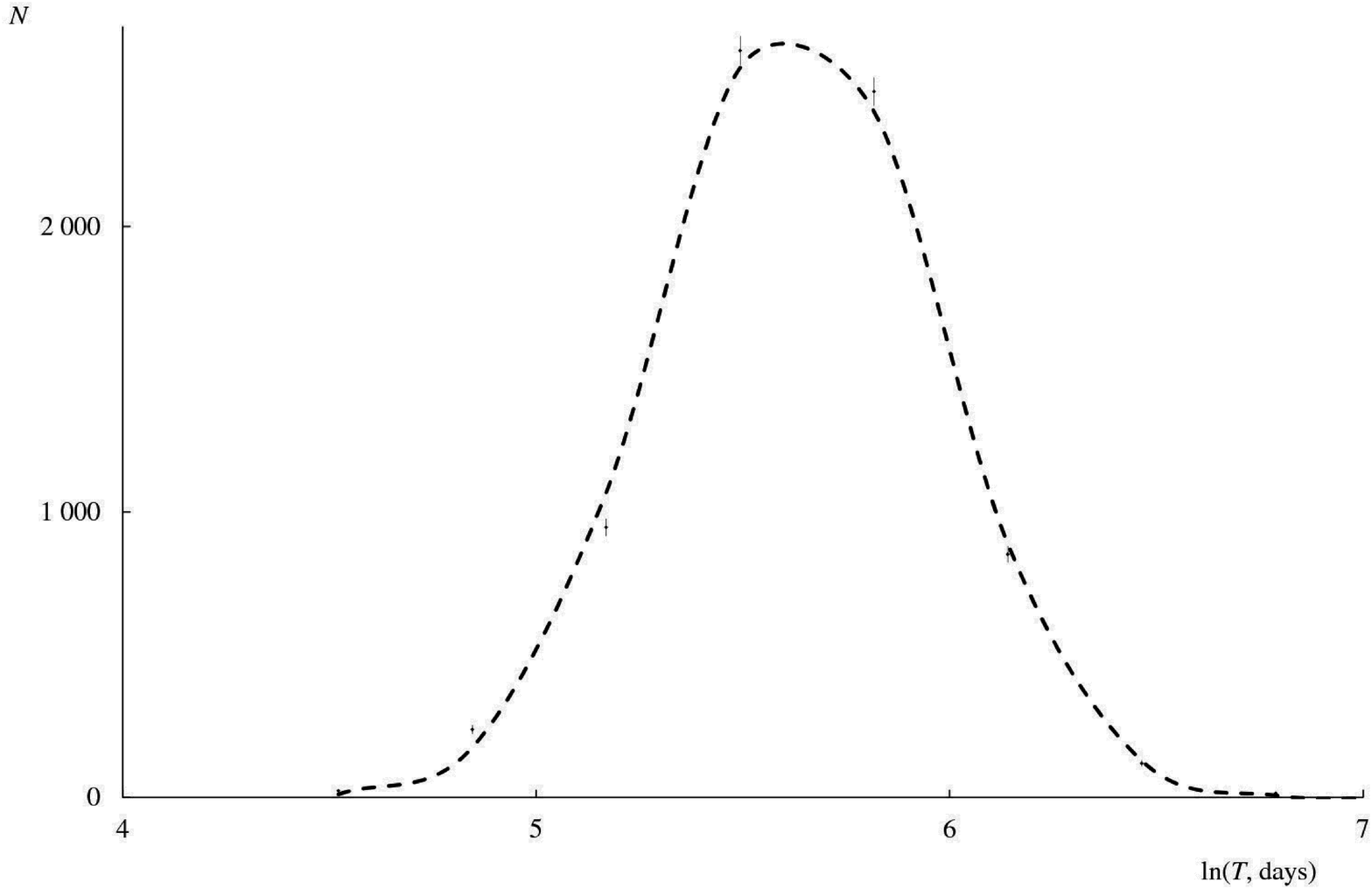}\\
(a)&(b)
\end{tabular}
\caption{Examples of log-period distributions for stars of type variability: (a) SRA (red giants of late spectral classes) and (b) M (red giants at the final stages of stellar evolution).}
\label{fig6}
\end{figure*}

The distribution of stars over log-periods corresponds to the normal distribution $N~\sim~\mathrm{exp}(-\frac{(\mathrm{ln}(T)-\mathrm{ln}(\bar{T}))^2}{2\sigma^2})$ with the parameters given in Table~\ref{tabl:2}. Stars of SRA (193 days), M (280 days) and SRC (372 days) variabilities (Table~\ref{tabl:2}), examples of the distributions of which are given in Figure~\ref{fig6} become closest to the period of 229 days obtained by the Fourier analysis of the change in the PCR spectrum by $E_0$ $\gamma$ index (Figure~\ref{fig4}).

\begin{table}[htbp]
\caption{Parameters of log-distributions of stars with SR, SRA, SRB, SRC, and M variabilities.}
  \centering
    \begin{tabular}[t]{|c|c|c|c|}
\hline    
Star variability & $\bar{T}$, days & $\sigma$ & $R_a^2$ \\
\hline
SRB & 105 & 0.75 & 0.97 \\
SR & 124 & 0.84 & 0.99 \\
SRA & 193 & 0.49 & 0.96 \\
M & 280 & 0.34 & 0.99 \\
SRC & 372 & 0.92 & 0.90 \\
\hline
    \end{tabular}
\label{tabl:2}    
\end{table}

It follows from Figure 6 that the number of Miras is much greater than that of other types of stars with similar periods. It can be assumed that the bump in the PCR spectrum by $E_0$ around $100~PeV$ is formed mainly by Miras.

\begin{figure}[h]
\centering
\includegraphics[width=\linewidth]{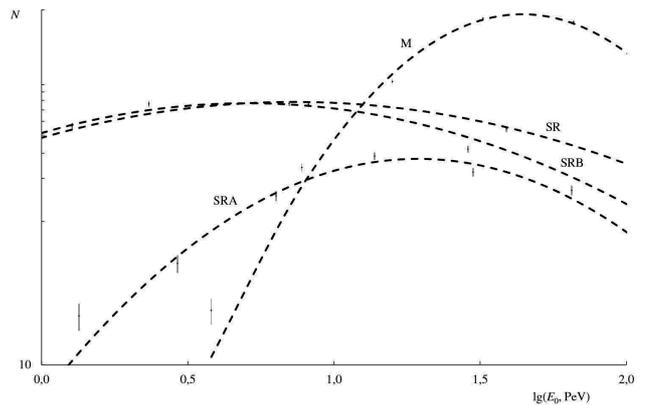}
\caption{Spectra of stars in log-energy variabilities SR, SRA, SRB, and M of spectral classes M (stars with neutral metals lines), C (carbon stars), and S (zirconium stars).}
\label{fig7}
\end{figure}

According to formula~(1) and with the parameters of normal distributions given in Table~\ref{tabl:2}, the spectra of stars by $E_0$ of various types of variability were obtained (Figure~\ref{fig7}).

It follows from Figure~\ref{fig7} that, as $E_0$ increases in the range $E_0~=~1-100~PeV$, stars at the final stages of stellar evolution start making decisive contribution to PCR, which makes the mass composition of PCR heavier. However, for each value of $E_0$, the PCR mass composition is determined by the dominant type of stars of a given period, which can lead to significant fluctuations in the fraction of different nuclei in the PCR mass composition with change on $E_0$.

\begin{figure}[h]
\centering
\includegraphics[width=\linewidth]{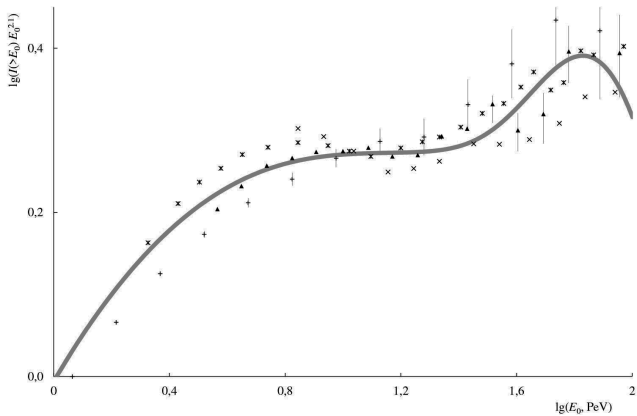}
\caption{Comparison of PCR spectra by $E_0$ obtained in the GAMMA (Armenia), \mbox{Tunka-133}, \mbox{Ice-Top}, and \mbox{KASCADE-Grande} experiments with the distribution of SR, SRA, SRB, SRC, and M type variability stars (gray curve).}
\label{fig8}
\end{figure}

PCR spectra by $E_0$ obtained in the GAMMA (Armenia), \mbox{Tunka-133}, \mbox{Ice-Top} and \mbox{KASCADE-Grande} experiments as compared with the spectra of dominant stars of SR, SRA, SRB, SRC and M variabilities obtained from approximations with parameters from Table~\ref{tabl:2} are shown in Figure~\ref{fig8}. The peak of the bump in the PCR spectrum obtained from these approximations is $E_0~\cong~67~PeV$.

However, it should be noted that, as follows from Figure~\ref{fig8}, the bump should be less pronounced and be at $E_0~<~67~PeV$. For example, the average period of SRC variability stars (super-giants) is 372 days, which, according to formula (1), gives the value lg$(E_0)~=~1.91 (81~PeV)$. This value of $E_0$ corresponds to the results of the \mbox{KASCADE-Grande} experiment. At the same time, the average period of Miras is 280 days, or lg$(E_0)~=~1.64 (44~PeV)$. Because since the number of observed Miras is an order of magnitude greater than that of stars with SRC variability, the localization of the bump maximum according to the \mbox{KASCADE-Grande} experiment at $E_0~=~80~PeV$ may be overestimated. It can also be assumed that the bump is formed both by Miras and by SRC super-giant stars.

The types of dominant stars for different $E_0$ are listed in Table~\ref{tabl:3}. With the increase in the mass of a star (system of stars) the period of the star and the maximum energy $E_0$ of PCR nuclei, provided by the acceleration mechanism in flares of this type of star increase.

\begin{table*}[htbp]
\caption{Types of dominant stars for different $E_0$ (EA are Algol sub-giants of late evolution, BY are draconian dwarfs (for example, the Sun~\cite{9}) of the Ke and Me spectral classes, RS are eruptive stars (a dwarf with a magnetically active sub-giant as secondary component) with Ca-II, H, and K spectra in emission).}
  \centering
    \begin{tabular}[t]{|c|c|c|c|}
\hline    
Energy, lg$(E_0, PeV)$ & Star period, days & \multicolumn{2}{c|}{Star type} \\
\hline
$<~0.1$ & $<~17$ & Sub-dwarfs, dwarfs & EA, BY, RS \\
\hline
$0.2-2$ & $23-74$ & Semi-regular giants & SR, SRB \\
\hline
$>~20$ & $>~176$ & Giants, super-giants & SR, SRB, SRC, M \\
\hline
$>~10^6$ & $>~29,000$ & Recurrent novae & NR \\
\hline         
    \end{tabular}
\label{tabl:3}    
\end{table*}

\begin{figure*}[h]
\centering
\includegraphics[width=\linewidth]{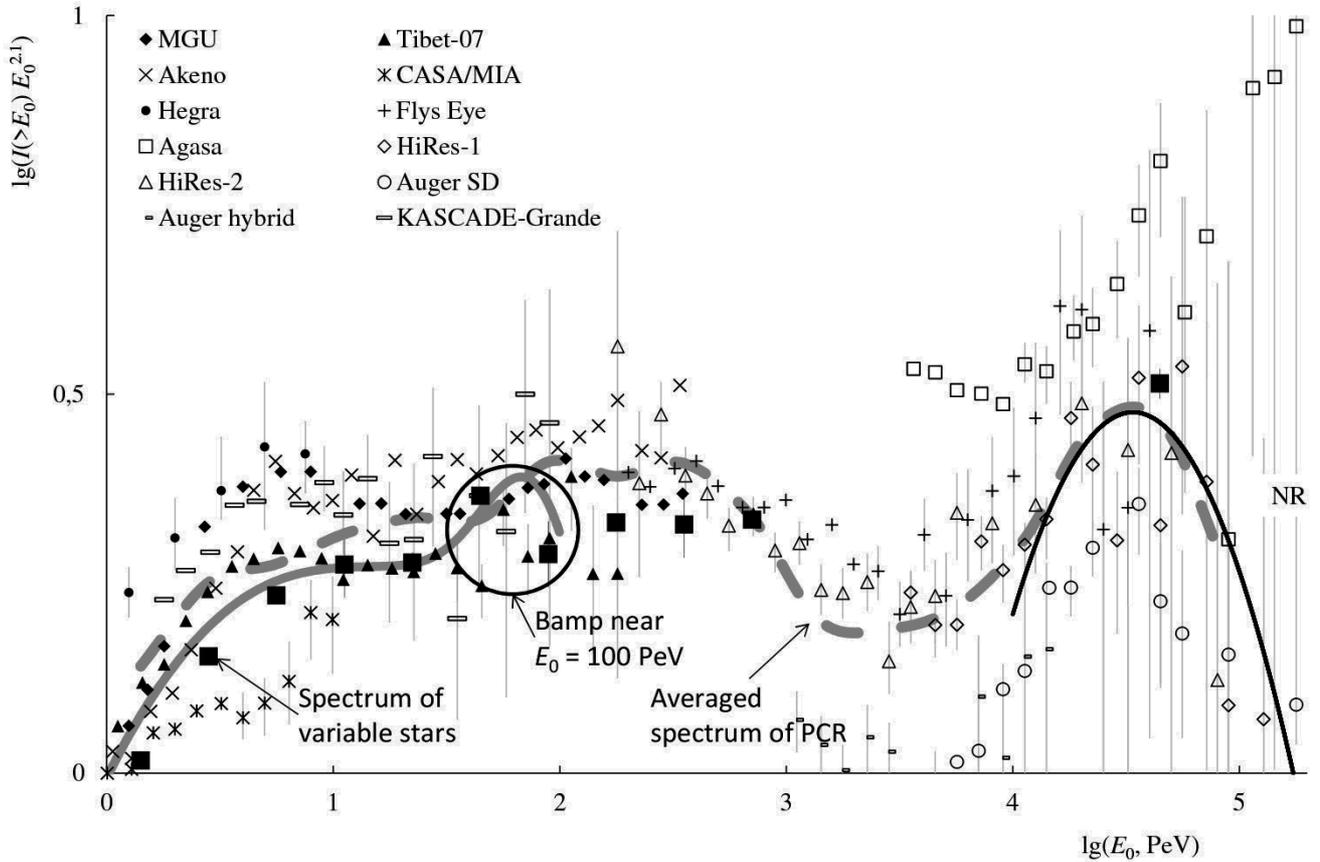}
\caption{Comparison of the PCR spectra for $E_0~>~1~PeV$ obtained in a number of experiments~\cite{10,11,12,13,14,15,16,17,18,19,20} with the spectrum of variable stars. A bump near $E_0~=~100~PeV$ is provided by the stars of the Miras and SRC variabilities. In the range $E_0~=~200~PeV~- 3.5~EeV$ (lg$(E_0,~PeV)~=~2.3-3.5)$, there are no stars with the corresponding periods (not registered), which determines the minimum in the PCR spectrum by $E_0$ in this range.}
\label{fig9}
\end{figure*}

Figure~\ref{fig9} compares the PCR spectra by $E_0$ obtained in a number of experiments with the spectrum of variable stars. CR acceleration to ultra-high energies occurs in explosive and nova-like stars, for example, in recurrent novae, in which the maximum period has been registered, providing the maximum $E_0~\sim~1-2~ZeV$. In the range $E_0~=~200~PeV~-~3.5~EeV$, no stars of variable types have been registered (see Figure~9): ZAND type of variability corresponds to the average period $T~=~553$ days, or according to formula~(1) $E_0~=~200~PeV$, followed by stars of $N$ variability with $T~=~2000$ days, or $E_0~=~3.5~EeV$.

An example of a binary star system where acceleration to ultra-high energies occurs can be an EA+SRC variability star with a recorded period of 7,430~days (20~years), which should provide a bump in the PCR spectrum at $E_0~=~60~EeV$ or lg$(E_0,~PeV)~=~4.80$. It can be a $\mu$~Cepheus type star (Herschel's garnet, a red super-giant at the last stage of stellar evolution with a He-C cycle) and an Algol $\beta$~Perseus type star.

An example of a triple star system where acceleration to ultra-high energies occurs can be a system of UGSU+E+ZZ variability stars with a recorded period of 11,900~days (33~years), which should provide a bump in the PCR spectrum at $E_0~=~180~EeV$ or lg$(E_0,~PeV)~=~5.26$. This may be a system of dwarf stars of the SU Ursa Major type (an eruptive dwarf) with super-maximal flares with amplitude of up to $2^m$. If this source is considered the only one providing the PCR flux at $E_0~=~180~EeV$, then the changes in the PCR flux at a given $E_0$ over 33~years should be significant, from the maximum to complete decay, which is observed in experiments.

\begin{figure}[h]
\centering
\includegraphics[width=\linewidth]{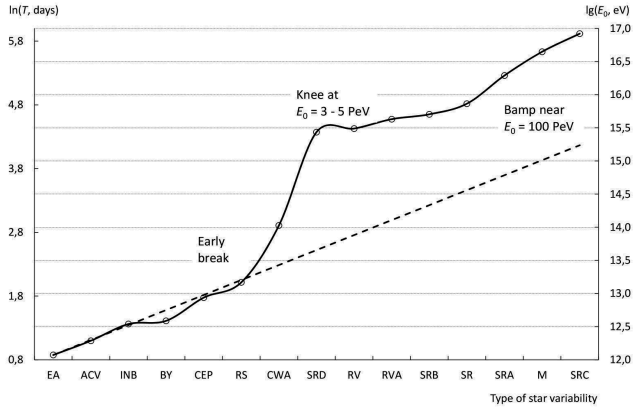}
\caption{Dependence of stellar periods and energy $E_0$ on the types of eclipsing variable stars.}
\label{fig10}
\end{figure}

Summarizing the results obtained in this study, let us once again consider the characteristics of variable stars from the GCVS catalog~\cite{5}. Let us sort the stars according to the increasing period and construct a double dependence of the log-period and $E_0$ on the types of eclipsing variable stars shown in Figure~\ref{fig10}. This spectrum is characterized by three main regions of irregularities with respect to linear dependence: starting from the RS variability stars, the so-called early knee in the PCR spectrum by $E_0$; starting from SRD type stars (giants and super-giants of spectral classes F, G, or K), there is a knee at $E_0~=~3-5~PeV$; starting with Miras,~- the so-called bump at $E_0$ about $100~PeV$. It should also be noted that, as follows from Figure~\ref{fig10}, after the steepening of the PCR spectrum by $E_0$ after the knee at $3-5~PeV$, the $\gamma$ index of the PCR spectrum by $E_0$ decreases again and becomes approximately the same as it was before the knee.

{\bf Conclusions.}
The relationship between the $E_0$ spectrum of PCR and the period spectrum of variable stars is shown. It is shown that the bump in the PCR spectrum at $E_0~=~3-5~PeV$ is explained by the change in the type of source stars in this range of $E_0$. The factor of PCR acceleration to high and ultra-high energies not only on supernova shock waves, but also in flares of giant and super-giant stars is responsible for the shape of the PCR spectrum by $E_0$.

\begin{enumerate}
\item PCR sources are variable stars of various types, which are at different stages of evolution, from sub-dwarfs to super-giants.
\item CR acceleration to high and super-high energies occurs not only on the shock waves of supernovae, but also in flares of giants and super-giants.
\item There is a direct relationship between the average period for a star of a given type of variability and the maximum PCR energy $E_0$ provided by the acceleration mechanisms on these stars.
\item A star of each type determines the $E_0$ PCR range with its flares. Each $E_0$ range corresponds to the PCR mass composition, which is determined by the type of the source star and which can change significantly with $E_0$.
\item The bump in the PCR spectrum by $E_0$ about $100~PeV$ is formed by giants and super-giants of M and SRC variability of late spectral classes. Other types of stars are responsible for other irregularities in the PCR spectrum by $E_0$.
\item The maximum energy $E_0$ of PCR is determined by recurrent novae and amounts to $1-2~ZeV$. It is also possible that there are longer-period types of stars that provide higher values of $E_0$.

\end{enumerate}


\begin{thebibliography}{99}
\bibitem{1}
Erlykin~A.~D., Puchkov~V.~S., Pyatovsky~S.~E. Change in the mass composition of primary cosmic radiation at energies in the range of $E_0~=~1-100~PeV$ according to data of the \mbox{KASCADE-Grande} experiment//Physics of Atomic Nuclei. - 2021. - Vol. 84. - No 3. - p. 279-286. - DOI: 10.1134/S1063778821030170

\bibitem{2}
T.~Antoni, W.~D.~Apel, F.~Badea, K.~Bekk, A.~Bercuci, H.~Blumer, H.~Bozdog, I.~M.~Brancus, C.~Buttner, A.~Chilingarian, K.~Daumiller, P.~Doll, J.~Engler, F.~Febler, H.~J.~Gils, R.~Glasstetter, et al., Nucl. Instrum. Methods Phys. Res., Sect. A 513, 490 (2003). - DOI: 10.1016/S0168-9002(03)02076-X

\bibitem{3}
A.~P.~Garyaka, R.~M.~Martirosov, S.~V.~Ter-Antonyan, A.~D.~Erlykin, N.~M.~Nikolskaya, Y.~A.~Gallant, L.~W.~Jones, and J.~Procureur, J. Phys. G: Nucl. Part. Phys. 35, 115201 (2008); arXiv: 0808.1421v1 [astro-ph]. - DOI: 10.1088/0954-3899/35/11/115201

\bibitem{4}
Apel~W., Arteaga~J.~C., et al. The \mbox{KASCADE-Grande} experiment//\mbox{KASCADE-Grande} Collaborations, Nucl. Instrum. Methods Phys. Res. A 620 (2-3) (2010), pp. 202-216. - DOI: 10.1016/j.nima.2010.03.147

\bibitem{5}
General Catalogue of Variable Stars//The Sternberg Astronomical Institute, The Institute of Astronomy of Russian Academy of Sciences. URL: http://www.sai.msu.su/gcvs/

\bibitem{6}
C.~Xiaodian, W.~Shu, D.~Licai, et al. The Zwicky Transient Facility Catalog of Periodic Variable Stars//The Astrophysical Journal Supplement Series, 249:18 (21pp), 2020 July. DOI - 10.3847/1538-4365/ab9cae

\bibitem{7}
S.~B.~Shaulov, V.~A.~Ryabov, A.~L.~Schepetov, S.~E.~Pyatovsky, et al. Strange quark matter and the astrophysical nature of anomalous effects in cosmic rays at energies of $1-100~PeV$//Letters to the Journal of Experimental and Theoretical Physics. 2022. 1-2(7). 116. с. 3-12. DOI - 10.31857/S1234567822130018

\bibitem{8}
V.~G.~Sinitsyna, V.~Yu.~Sinitsyna, Yu.~I.~Stozhkov. Red dwarf stars as a new source type of galactic cosmic rays//Astronomische Nachrichten. 2021. 342. 1-2. pp. 342-346. DOI - 10.1002/asna.202113931

\bibitem{9}
I.~Yu.~Alekseev. Statistics of BY Draconis Variables//Astron. Rep. 44, 696-700 (2000). DOI - 10.1134/1.1312966

\bibitem{10}
De~Mitri~I. on behalf of the ARGO-YBJ Collaboration. Measurement of the cosmic ray all-particle and light-component energy spectra with the ARGO-YBJ experiment//ISVHECRI 2014. - 18th International Symposium on Very High Energy Cosmic Ray Interactions, EPJ Web of Conferences. - 2015. - 99. p. 08003. - DOI: 10.1051/epjconf/20159908003

\bibitem{11}
Apel~W.~D., Arteaga-Velazquez~J.~C., et al. \mbox{KASCADE-Grande} measurements of energy spectra for elemental groups of cosmic rays//Astropart. Phys. - 2013. - 47. - pp.54-66. - DOI: 10.1016/j.astropartphys.2013.06.004

\bibitem{12}
Apel~W.~D., Arteaga~J.~C., et al. Energy spectra of elemental groups of cosmic rays: Update on the KASCADE unfolding analysis//Astropart. Phys. 31 (2009), 2, pp. 86-91. - DOI: 10.1016/j.astropartphys.2008.11.008

\bibitem{13}
Schoo~S., Kang~D., et al. - \mbox{KASCADE-Grande} Collaboration. A new analysis of the combined data from both KASCADE and \mbox{KASCADE-Grande}//Proceedings of the 35th International Cosmic Ray Conference. - 2017. - PoS(ICRC2017). - 339

\bibitem{14}
Kampert~K.-H., Unger~M. Measurements of the cosmic ray composition with air shower experiments//Astropart. Phys. - 2012. - 35. - pp.660-678. - DOI: 10.1016/j.astropartphys.2012.02.004

\bibitem{15}
Budnev~N., Astapov~I., et al. The TAIGA experiment - a hybrid detector for very high energy gamma-ray astronomy and cosmic ray physics in the Tunka valley//Proceedings of the 35th International Cosmic Ray Conference. - 2017. - PoS(ICRC2017). - 768

\bibitem{16}
Fedorov~O., Bezyazeekov~P.~A., et al. - Tunka-Rex Collaboration. Detector efficiency and exposure of Tunka-Rex for cosmic-ray air showers//Proceedings of the 35th International Cosmic Ray Conference. - 2017. - PoS(ICRC2017). - 387

\bibitem{17}
Sveshnikova~L., Astapov~I., et al. Search for gamma-ray emission above 50 TeV from Crab Nebula with the TAIGA detector//Proceedings of the 35th International Cosmic Ray Conference. - 2017. - PoS(ICRC2017). - 677

\bibitem{18}
Porelli~A., Wischnewski~R., et al. TAIGA-HiSCORE detection of the CATS-LIDAR on the ISS as fast moving point source//Proceedings of the 35th International Cosmic Ray Conference. - 2017. - PoS(ICRC2017). - 754

\bibitem{19}
Postnikov~E., Astapov~I., et al. Commissioning the joint operation of the wide angle timing HiSCORE Cherenkov array with the first IACT of the TAIGA experiment//Proceedings of the 35th International Cosmic Ray Conference. - 2017. - PoS(ICRC2017). - 756

\bibitem{20}
Kopper~C. - IceCube Collaboration. Observation of Astrophysical Neutrinos in Six Years of IceCube Data//Proceedings of the 35th International Cosmic Ray Conference. - 2017. - PoS(ICRC2017). - 981

\end{thebibliography}
\end{document}